\documentclass[12pt]{article}
\usepackage[dvips]{graphicx}
\def\be{\begin{equation}}
\def\ee{\end{equation}}
\def\ber{\begin{eqnarray}}
\def\eer{\end{eqnarray}}
\def\dint{\mathop{\intop\kern-0.5em\intop}}
\def\ovc#1{\displaystyle\mathop{#1}^{\kern0.2em\circ}}

\begin{document}

\begin{center}
{\bf\Large Comment on Andr\'e Martin: \\[1ex] "Can one improve
the Froissart Bound?"}\\

\vspace{4mm}

{\bf\large Andrei A. Arkhipov}\\[1ex]
{\it State Research Center ``Institute for High Energy Physics" \\
 142280 Protvino, Moscow Region, Russia}\\
\end{center}

\begin{abstract}
In this note my personal point of view on the question brought up for
a discussion at the Conference "Diffraction 2008" by Andr\'e Martin
has been presented.
\end{abstract}



\section*{}

\noindent

In recent article \cite{1}, which is really a typescript of the talk
given at the Conference "Diffraction 2008" held at
Lalonde-les-Maures, France, September 2008, Andr\'e Martin discussed
a possibility of improving the famous Froissart bound \cite{2} either
{\it qualitatively} or {\it quantitatively}. Qualitative improving
concerns $\ln^2s$ - dependence of the Froissart bound, it means the
replacing the exponent of the logarithm by an exponent smaller than
2. Quantitative improving concerns the constant staying in the
Froissart bound for the total cross section, and this is an attempt
to replace the constant by a smaller one. In this note our personal
point of view on the problems under discussion has been presented.

First of all, I would like to note that $\ln^2s$ - dependence of the
Froissart bound cannot be improved. This is quite clear if we rewrite
the Froissart bound for the total cross section of two-body reaction
$a+b\rightarrow a+b$ in the form \cite{3}
\begin{equation}
\sigma_{ab}^{tot}(s)<4\pi R_2^2(s).\label{1}
\end{equation}
Here the quantity $R_2(s)$ has a strong mathematical definition with
a clear and transparent physical meaning; see details in Ref.
\cite{3} and references therein
\begin{eqnarray}
R_2(s)&\stackrel{def}{=}&\frac{L}{|\bf q|} =
\frac{2\sqrt{s}\ln{\tilde
P}_2(s)}{\sqrt{2\epsilon(s)\lambda(s,m_a^2,m_b^2)}} = \frac{\ln
{\tilde P}_2(s)}{\sqrt{t_0}}\\
&\simeq& \frac{9}{4\sqrt{t_0}}\ln
(s/s_0) = \frac{9}{8m_\pi}\ln (s/s_0),\quad s \gg s_0, \quad
(t_0\equiv 4m_{\pi}^2). \label{2}
\end{eqnarray}
The pion mass $m_{\pi}$ in R.H.S. of Eq. (\ref{2}) appears from the
nearest $t$-channel threshold, $s_0$ is a determinative scale usually
extracted from a fit to experimental data. The quantity $R_2(s)$ is
named as the effective radius of two-body forces, and it is simply
related with the experimentally measurable quantity which is the
slope of diffraction cone $B_{el}(s)$ in elastic forward scattering
for the two-body reaction
\begin{equation}\label{3}
B_{el}(s) = \frac{1}{2}R_2^2(s).
\end{equation}
It should especially be emphasized that the quantity $R_2(s)$
accumulates all information concerning polynomial boundedness  and
analyticity of the two-body reaction amplitude in a topological
product of complex $s$-plane with the cuts ($s_{thr}\leq s
\leq\infty,\, u_{thr}\leq u \leq\infty$) except for possible fixed
poles and circle $|t|\leq t_0$ in complex $t$-plane, where $s$, $t$,
$u$ are Mandelstam variables. That analyticity is proved in the
framework of axiomatic Quantum Field Theory, and this is enough to
save and extend the fundamental Froissart result previously obtained
at a more restricted Mandelstam analyticity. Much technical work was
done by G.~Sommer, A.~Martin and others to extend the Froissart
theorem to the domain of axiomatic Quantum Field Theory. However, in
our opinion, the corner stone in that extension has to be referred to
Harry Lehmann \cite{4} who proved that two-body elastic scattering
amplitude is analytic function of $\cos\theta$, regular inside an
ellipse in complex $\cos\theta$-plane with center at the origin. The
fundamental Jost-Lehmann-Dyson representation - brilliant
quintessence of general principles in the theory of quantized fields
- especially Dyson's theorem for a representation of causal
commutators in local Quantum Field Theory \cite{5,6,7} and not more
have been used by Harry Lehmann. From the fundamental result of Harry
Lehmann it follows that the partial wave expansions which define
physical scattering amplitudes continue to converge for complex
values of the scattering angle, and define uniquely the amplitudes
appearing in the nonphysical region of non-forward dispersion
relations. In fact, expansions converge for all values of momentum
transfer for which dispersion relations have been proved. It would be
interested to remember at this place the real story which I have
heard from my friends. It was in the middle of 1950th at the
Conference held in Dubna, where Lehmann reported his above mentioned
result in the first time. The attended there Landau has argued during
the talk that the Lehmann result was wrong because in his opinion
two-body elastic scattering amplitude must be analytic function in
the whole complex plane. Harry Lehmann wittily replied: Landau is a
big man, so he needs analyticity of the amplitude in the whole
complex plane, but I am a small man, and analyticity of the amplitude
in the ellipse is enough for me.\footnote{Probably, it should to be
added here another real story. Close on the same time Landau has
argued that Bogoljubov's article on superconductivity in the metals
was also wrong even though the article was really impeccable
mathematically and well grounded physically. Landau was informed that
his disciple Migdal wrote and prepared for publication the paper
where he shown that the superconductivity in the metals was
impossible. The Migdal's paper was already accepted for publication
in the Soviet Journal JETP. However, just  in that time the famous
article of J.~Bardeen, L.N.~Cooper and J.R.~Schriffer {\it Theory of
superconductivity} has been appeared in the Physical Review ({\bf
108}, 1175 (1957)). The Migdal's paper was promptly withdrawn from
publication, and Landau was deeply disappointed.}

In addition, it turns out that the Froissart bound with $\ln^2s$ -
dependence is saved even in a special class of nonlocal Quantum Field
Theories where the amplitude may exponentially increase in the
complex momentum square $p^2$-space (the so-called essentially
nonlocalizable field theories): ${\cal A} \sim \exp(l^{2}p^{2})$,
where $l$ determines the scale of nonlocal interaction. This was
shown in Ref. \cite{8}. Axiomatic status of nonlocal Quantum Field
Theory can be found in Refs. \cite{9,10,11,12} of Russian schools
directed by Vladimir Fainberg and Garry Efimov independently.

All of that means that the Froissart bound represents a physically
tangible consequence from abstract mathematical structures given by
general axioms in the theory of quantized fields even nonlocal ones.
That is why, the Froissart bound is often considered as intrinsic
property of the theory of quantized fields.

In our opinion, the bound (\ref{1}) represents the most rigorous
mathematical formulation of the holographic principle \cite{13} which
is widely discussed in the recent literature. Thus the holographic
principle has been incorporated in the general scheme of axiomatic
Quantum Field Theory and resulted from the general principles of the
theory of quantized fields \cite{3}.

From the Froissart bound in the case of the two-body forces saturated
unitarity one obtains
\begin{equation}\label{4}
\sigma_{ab}^{tot}(s) = 4\pi R_2^2(s)\simeq C_{ab}\ln^2(s/s_0), \quad
s\rightarrow\infty,
\end{equation}
where
\begin{equation}\label{5}
C_{ab}=\frac{4\pi\cdot 81}{64\,m_{\pi}^2}=\frac{15.9}{m_{\pi}^2}\cong
339\, \mbox{mb}.
\end{equation}
Certainly, the value 339 mb for the constant $C_{ab}$ is too large to
fit to available experimental data. However, it is quite clear, this
only means that the two-body forces  do not saturate unitarity in the
range of reachable energies at now working accelerators, and Eq.
(\ref{1}) really represents the bound only. Just in case, I would
like in this note to draw attention to elegant way for
structurization of the constant $C_{ab}$ in R.H.S. of Eq. (\ref{4})
if we will take into account not only two-particle but three-particle
unitarity as well.

Careful study of the problem of scattering from the deuteron with a
complicated analytical work and tedious calculations (see details in
recent review article \cite{14} and references therein) allowed us to
derive the theoretical formula describing the global properties of
(anti)proton-proton total cross sections which has the following
structure \cite{15}
\begin{equation}\label{6}
\sigma_{(\bar p)pp}^{tot}(s) = [1+\chi_{(\bar
p)pp}(s)]\sigma_{asmpt}^{tot}(s).
\end{equation}
The structure given by Eq. (\ref{6}) has been appeared as consistency
condition to fulfil the unitarity requirements in two-particle and
three-particle sectors simultaneously. In according to this structure
the total cross section is represented in a factorized form. The
first factor is responsible for the behavior of total cross section
at low energies with the universal energy dependence at elastic
threshold, it has a complicated resonance structure, and $\chi(s)$
tends to zero at $s\rightarrow \infty$. The other factor describes
high energy asymptotic of total cross section, and it has the
universal energy dependence predicted by the general theorems in
Quantum Field Theory. For this factor one obtains
\begin{eqnarray}
\sigma_{asmpt}^{tot}(s) &=& 2\pi B_{el}(s) + 4\pi (1-\beta)B_{sd}(s)
= 2\pi B_{el}(s)[1 + 2\gamma(1-\beta)]\\[1ex]
&=& \pi R_2^2(s) + 2\pi (1-\beta)R_3^2(s)\, = \,\pi R_2^2(s)[1 +
2\gamma(1-\beta)], \label{7}
\end{eqnarray}
where $B_{sd}(s)$ is the slope of diffraction cone for inclusive
diffraction dissociation processes, $R_3(s)$ is the effective radius
of three-nucleon forces related to the slope $B_{sd}(s)$ in the same
way $B_{sd}(s) = R_3^2(s)/2$ as the effective radius of two-nucleon
forces is related to the slope $B_{el}(s)$ of diffraction cone in
elastic scattering processes, $\beta$ is slowly energy dependent
dimensionless quantity from interval $0\leq \beta \leq 1/4$, $\beta$
tends to 1/4 at $s\rightarrow \infty$ and $\beta \ll 1$ up to LHC
energies, and  $\gamma = B_{sd}(s)/B_{el}(s) = R_3^2(s)/R_2^2(s)$
obviously. From the Froissart bound it follows $\gamma < 2$.

It is a non-trivial fact that the constant in R.H.S. of Eq.
(\ref{7}), staying in front of effective radius of two-nucleon
forces, is 4 times smaller than the constant in the Froissart bound.
But this is too small to correspond to the experimental data if we
use the experimental data on $B_{el}(s)$. The second term in R.H.S.
of Eq. (\ref{7}) fills an emerged gap.

Using formula (\ref{6}), we have made the global fit \cite{15} to the
experimental data on $pp$ and $p\bar p$ total cross sections in the
whole range of energies available up today and found a very good
correspondence of the theoretical formula to the existing
experimental data obtained at the accelerators. Moreover, it was
shown a very good correspondence of the theory to all existing cosmic
ray experimental data as well \cite{16}. The predicted values for
$\sigma^{tot}_{pp}$ obtained from theoretical description of all
existing accelerators data are completely compatible with the values
obtained from cosmic ray experiments. At the LHC we predict
\begin{equation}\label{8}
\sigma_{pp}^{tot}(\sqrt{s}=14\,\mbox{TeV}) = 116.53 \pm
3.52\,\mbox{mb}.
\end{equation}
It should to be compared to the best even though very crude estimate
based on {\it Pomeron Physics and QCD} \cite{17}
\begin{equation}\label{9}
\sigma^{LHC} = 125 \pm 35\,\mbox{mb},
\end{equation}
presented by Peter Landshoff at the same Conference "Diffraction
2008" \cite{18}.\footnote{It should to be emphasized that there is no
common agreement among even physicists  fully or blindly accepted the
Pomeron phenomenology. Our criticism of the Pomeron phenomenology can
be found in \cite{19}; see also \cite{15}.}

In the case of the two-body forces saturated the Froissart bound,
taking into account that $\sigma_{el}=\sigma_{tot}^2/16\pi
B_{el}\,(\rho_{el} = 0)$, and $B_{el}=R_2^2/2$, one obtains
\begin{equation}\label{10}
\sigma_{ab}^{tot}(s) = 4\pi R_2^2(s)\quad \Rightarrow \quad
\sigma_{ab}^{el}(s)=\frac{1}{2}\,\sigma_{ab}^{tot}(s),\quad
s\rightarrow\infty.
\end{equation}
Thus we come to the following statement: {\it The two-body forces
saturated the Froissart bound saturate the Pumplin bound as well}.

To conclude, in our opinion, there is no any necessity in improving
the fundamental Froissart theorem. However, we can generalize the
theorem though to the case of multiparticle interaction using the
multidimensional space picture; see details in \cite{3} and
references therein. The general principles in the theory of quantized
fields together with constructive dynamic apparatus of Bethe-Salpeter
type formalism in QFT provide the clear answers to the question
brought up for a discussion in \cite{1}. The very core of our
approach to the problem in understanding the total cross section
consists in that we have to go beyond the study of a dynamics of pure
two-body reaction and consider the dynamics of a more complex
three-body system. In this way we have established that the
relativistic dynamics of three-body system with a necessity contains
the fundamental three-body forces which are, in one's turn, related
to the dynamics of one-particle inclusive reactions. Moreover, as was
demonstrated above the effective radii of two- and three-body forces
being the characteristics of elastic and inelastic interactions in
two-body subsystems have been combined in a special form determining
the nontrivial dynamical structure for the total cross-section
clearly confirmed by the experimental data.

Finally, paraphrasing Harry Lehmann, I would say that the Froissart
bound is quite enough for me, convenient for me and suit me.

\section*{Acknowledgements}
I'd like to thank Andr\'e Martin for his continuous interest to the
Froissart theorem. No doubt, the Froissart theorem is the greatest
theorem in theoretical particle physics of the second half of XX
century. I am sincerely grateful to Marcel Froissart for his kindest
letter with friendly encouragement as well.

\end{document}